\documentclass[11pt]{article}

\usepackage{mystyle}
\usepackage{contributors}
\usepackage{mymath}

\title{Noise Effects on Ordinal Pattern Statistics via Majorization}

\date{}

\begin{document}
\maketitle

% \hfill \break
% \thanks
% \newpage

\begin{abstract}
The Bandt-Pompe permutation entropy framework, alongside the complexity-entropy causality plane, has become a standard tool for characterizing the dynamical properties of time series. 
However, observational noise distorts ordinal pattern probability distributions in ways that can systematically misplace time series within the causality plane, compromising dynamical classification. 
This effect is particularly relevant for geophysical signals, which are typically poorly and irregularly sampled, and have a low signal-to-noise level.
In this work, we characterize the distortions on ordinal pattern statistics using the formalism of majorization. 
We provide theoretical results and propose corrective strategies that restore discriminability under realistic measurement conditions. 
To achieve this, we introduce methodology that allows the characterization of noisy dynamical series and further allows the quantification of observational noise without the need of a fitting procedure. 
Finally, we illustrate our methodology by analyzing paleomagnetic records to determine if the geological evolution of the Earth dipole is better described by a stochastic or chaotic system.
\end{abstract}

\section{Introduction}
The characterization of a given time series is an important problem in data analysis across scientific domains, providing us with a better understanding of the processes and mechanisms governing the system under study.
Since its introduction by \textcite{bandt2002permutation-d0b}, the use of \textit{ordinal patterns} has proven to be a useful tool for addressing this task.
Instead of looking at the complete structure and complexity of a time series, ordinal patterns reduce the study of a time series to its numerical ordering using the embedding approach introduced by \textcite{Takens_1981}, thereby preserving the temporal/sequential structure of the data while discarding its amplitude information.
This property is why ordinal patterns have proven so effective at distinguishing chaotic from stochastic dynamics, even when these two share so much in common (see \textcite{zanin2021ordinal-4a8} for a comprehensive review).

A key advantage of ordinal pattern analysis is that it is effective, simple to compute and prototype, and does not require model assumptions nor an inference/filtering procedure for the underlying time series \cite{Ricci_Perinelli_2022}.
Within this framework, it is common to reduce the information in the ordinal pattern distribution to an information-theory metric \cite{Zunino_Soriano_Rosso_2012}.
Common metrics computed on the ordinal patterns include entropies and complexity metrics.
The former, known as the permutation entropy, quantifies the degree of ordering in the time series and has the appealing property that it converges to the Kolmogorov-Sinai entropy \cite{bandt2002permutation-d0b}.
On the other hand, a second metric of complexity has been introduced by \textcite{rosso2007distinguishing-a8b} to complement the permutation entropy, as the former alone may not be sufficient to characterize the nature of a time series.
These two can further be combined in a two-dimensional diagram to result in the \textit{complexity-entropy causality plane} \cite{rosso2007distinguishing-a8b}.
Within the complexity-entropy causality plane, different types of time series (e.g., chaotic and stochastic) display a localization behavior that allows their classification.
Other analyses commonly performed on top of ordinal patterns include the characterization of forbidden patterns \cite{Amigo_Kocarev_Szczepanski_2006}, transitions between consecutive ordinal patterns \cite{Small_2013}, and multi-scale complexity-entropy planes \cite{Zunino_Soriano_Rosso_2012}.
See \textcite{ribeiro2017characterizing-bfd} for a comprehensive review on this topic.

Although all these properties make ordinal patterns a natural tool in the scientist's toolkit, their use in real-world applications is sometimes limited by the negative effects of observational noise and poorly sampled time series.
This limitation represents a barrier in utilizing ordinal pattern statistics in the natural and social sciences, where observations are usually sparse and with a low signal-to-noise ratio.
In these fields, we typically rely on short and noisy time series, where, for example, uncertainty quantification in the ordinal patterns plays a critical role \cite{Ricci_Perinelli_2022}.
Observational noise can significantly affect the characterization of chaotic and stochastic time series \cite{Lei_Meng_2008}, and can further change the values of permutation entropy and complexity \cite{ricci2021permutation-7d6, Zunino_Soriano_Rosso_2012, Porta_Bari_Marchi_Maria_Castiglioni_Rienzo_Guzzetti_Cividjian_Quintin_2015}.
While the robustness of ordinal pattern statistics to observational noise has been discussed in the literature, existing results are typically limited to small observational noise \cite{bandt2002permutation-d0b, ricci2021permutation-7d6, amigo2010permutation, Politi_Ricci_2025} or for long enough time series \cite{zanin2021ordinal-4a8}.

In this work, we address the problem of characterizing ordinal patterns in the presence of observational noise via the notion of \textit{majorization}.
Majorization between probability vectors offers a more localized view of the ordinal pattern distribution than information-based metrics — in line with the direction recently advocated in \textcite{zanin2021ordinal-4a8} — and this additional resolution is what allows us to characterize the effect of noise without requiring a full model of the dynamics.
Our central finding is that observational noise generates ordinal pattern probabilities that are majorized by the noiseless distribution, systematically shifting the permutation distribution toward uniformity and moving any time series to a predictable region of the causality plane.
Building on this, we make three main contributions:
\begin{enumerate}[label=(\roman*)]
    \item A formal proof of the majorization relation under certain observational noise models, and a numerical demonstration for more generic noise models.
    \item A hypothesis test that determines whether an observed distribution is consistent with a reference distribution, adapted to the finite-sample regime.
    \item An application to paleomagnetic records, where we use this framework to characterize the evolution of Earth's magnetic dipole.
\end{enumerate}

We apply the methods developed in this work to a real-world geophysical problem: determining whether the changes in the intensity of the dipole component of Earth's magnetic field over the past few million years are better described by chaotic or stochastic dynamics.
This question is of particular interest in the paleomagnetic literature, as both chaotic and stochastic reduced-order models are consistent with paleomagnetic observations, such as the reversal rate, the distribution of reversal intervals, and the statistics of dipole intensity fluctuations \cite{Gwirtz_Morzfeld_Fournier_Hulot_2020, Gwirtz_Davis_Morzfeld_Constable_Fournier_Hulot_2022}.
This problem is particularly suitable for the methodology introduced in this work, as paleomagnetic observations are good candidates for sparse and noisy observations.
We hope this example serves as a motivating demonstration of the potential of using ordinal patterns in other fields characterized by imperfect observations.

\section{Methods}
The framework of ordinal patterns to characterize discrete time series was introduced by the seminal work of \textcite{bandt2002permutation-d0b}.
Given a natural number $d$ known as the \textit{embedding dimension}, and a natural number $\tau$ known as the \textit{embedding delay}, the first step of the proposal is to use the embedding approach introduced by \textcite{Takens_1981} to transform the original time series $x_1, x_2, \ldots, x_N$, with $N$ the total length, into a collection of tuples $(z^i_1, z^i_2, \ldots, z_d^i) = (x_i, x_{i + \tau}, \ldots, x_{i + (d-1) \tau}) \in \R^d$, with $i = 1, 2, \ldots, N - (d-1)\tau$.
We then associate each tuple with an \textit{ordinal pattern} $\pi \in S_d$, with $S_d$ the symmetric group of degree $d$ composed of all $D = d!$ possible permutations, corresponding to the unique bijection $\pi : \{ 1, 2, \ldots, d\} \mapsto \{ 1, 2, \ldots, d\}$ such that $z_{\pi(1)}^i \leq z_{\pi(2)}^i \leq \ldots \leq z_{\pi(d)}^i$.
Finally, we define the probability $p(\pi)$ over $S_d$ as the relative frequency of each permutation $\pi$ in the set of all observed ordinal patterns.

There are a few considerations to address when dealing with ordinal pattern statistics.
First of all, edge cases, such as equal values in the time series and non-homogeneous sampling, can be easily addressed \cite{Riedl_Muller_Wessel_2013,zanin2021ordinal-4a8}.
More critically, the choice of the embedding parameters $\tau$ and $d$ have a significant impact on the ordinal pattern distribution.
The choice of $d$ further depends on the length of the time series, with $N \gg D$ a necessary condition to obtain reliable statistics, which in most time series leads to a maximum of $d \leq 7$ \cite{Riedl_Muller_Wessel_2013}.
Common choices include $d = 3, 4, 5, 6, 7$ \cite{bandt2002permutation-d0b}.
On the other hand, the delay parameter $\tau$ is associated with the scale at which we are characterizing the time series.
Short values of $\tau$ in the presence of noise can lead to a redundancy effect, while large values of $\tau$ can remove all relevant information of the time series \cite{Casdagli_Eubank_Farmer_Gibson_1991, Rosenstein_Collins_Luca_1994, Micco_Fernandez_Larrondo_Plastino_Rosso_2012}.
In many real-world applications, it is common to consider ordinal pattern statistics over a range of values of $\tau$ to capture different regimes (e.g., \textcite{Cencini_Falcioni_Olbrich_Kantz_Vulpiani_2000}).
Extensions to continuous time series $\{ x_t \}_{t \in \R}$ can also be considered.
In this case, the embedding delay $\tau$ is a positive real value and we apply the same procedure as in the discrete case over the set of tuples $(x_{t_0}, x_{t_0 + \tau}, \ldots, x_{t_0 + (d-1)\tau})$.

In real-world applications, one rarely has access to the noiseless state of the time series under study.
Instead, one observes an additive noise-corrupted version
\begin{equation}
    y_i = x_i + \varepsilon_i, \quad i = 1, 2, \ldots, N,
\end{equation}
where $\varepsilon_i$ denotes the observational noise \cite{brockwell1991timeseries, shumway2025timeseries, amigo2010permutation}.
A standard assumption in time series analysis is to model the observational noise as a zero-mean Gaussian process, $\varepsilon_i \sim \mathcal{GP}(0, k)$, where the covariance structure is fully characterized by a kernel function $k(i, j)$.
Common choices include white noise, $k(i, j) = \sigma^2 \delta_{i, j}$, and the exponentially decaying kernel, $k(i, j) = \sigma^2 \exp(-\theta\lvert i - j\rvert)$, which introduces temporal correlations controlled by the length-scale parameter $\theta > 0$.

\subsection{Ordinal Majorization}

Majorization is a partial order on the space of probability vectors that formalizes the notion of one distribution being more disordered than another \cite{Bickel}, with prominent applications in different areas of mathematics \cite{Bickel}, quantum information theory \cite{Nielsen1999, Horodecki2013}, and economics \cite{lorenz1905methods, ATKINSON1970244}, among others.
Given two probability vectors $p, q \in \mathbb{R}^D$, we say that $p$ is majorized by $q$, written $p \prec q$, if the partial sums of their decreasingly sorted components $p^{\downarrow}$ and $q^{\downarrow}$ satisfy 
\begin{equation}
    \sum_{j=1}^{k} p_j^{\downarrow} \leq \sum_{j=1}^{k} q_j^{\downarrow}
    \quad
    \forall k = 1, \ldots, D.
\end{equation}
An equivalent and more intuitive characterization states that $p \prec q$ if and only if $p$ can be written as a convex combination of permutations of $q$, making precise the sense in which $p$ is a more mixed version of $q$ \cite{Bickel}.

The following result shows that, under certain conditions, the distribution of the ordinal patterns obtained from the noisy time series $\tilde p$ is majorized by its noiseless version $p$.
This result imposes stronger constraints in ordinal pattern statistics, as it allows us to determine whether a noisy time series can be obtained from a reference noiseless time series, independently of the stochastic or chaotic nature of the original time series.
We present our result in the case of continuous and stationary processes, although this naturally also applies to the case of discrete time series.

\begin{theorem}
Let $\{x_t\} \subset \R$ be a strictly stationary process such that the distribution of the embedding $(x_t, x_{t+\tau}, \ldots, x_{t + (d-1)\tau}) \in \R^d$ is absolutely continuous with respect to the Lebesgue measure on $\mathbb{R}^d$ for any choice of $d \in \N$ and $\tau \in \R$.
Assume that the distribution of the sorted gaps between consecutive points $\Delta = (\Delta_1, \ldots, \Delta_{d-1}) \in \R^{d-1}$, with $\Delta_{i} = x_{\pi(i+1)} - x_{\pi(i)}$, is independent of the permutation order $\pi$.
Let $y_t = x_t + \varepsilon_t$, where $\varepsilon_t$ is a noise term such that the conditional joint distribution of the embedding $(\varepsilon_t, \varepsilon_{t + \tau}, \ldots, \varepsilon_{t + (d-1)\tau}) \in \R^d$ given the unsigned gaps $\Delta$ and the pattern $\pi$ associated to $(x_t, x_{t+\tau}, \ldots, x_{t + (d-1)\tau})$ is invariant under permutation of its components.
Let $p$ and $\widetilde{p}$ denote the ordinal pattern distributions associated with $\{x_t\}$ and $\{y_t\}$, respectively.
Then $\widetilde{p} \prec p$.
\label{theorem:main}
\end{theorem}

\begin{proof}
See Appendix \ref{sec:appendix-proof}.
\end{proof}

The conditional permutation-invariance assumption on the embedding of $\varepsilon_t$ is strictly weaker than independence-from-$\{x_t\}$ together with marginal exchangeability, and it covers a number of noise models of practical interest.
In particular, this includes the case of independent identically distributed noise, where $\varepsilon_t \sim F$ with $F$ any absolutely continuous distribution.
In the case of Gaussian noise, the most general noise model covered is
\begin{equation}
    \varepsilon_t, \varepsilon_{t + \tau}, \ldots, \varepsilon_{t + (d-1)\tau} \mid \Delta \;\sim\; \mathcal{N}\!\left(0,\; \Sigma(\Delta)\right),
    \qquad
    \Sigma(\Delta)_{ij} 
    = 
    \begin{cases}
    \sigma(\Delta) & i = j \\
    \rho(\Delta) & i \neq j \\
    \end{cases}
\end{equation}
with $| \rho(\Delta) | \leq \sigma(\Delta)$.
The covariance is equicorrelated for each realization of $\Delta$, but its scale and correlation strength may depend on the local geometry of the signal.
This includes the common case of independent Gaussian noise with $\rho(\Delta) = 0$ and the case of equicorrelated Gaussian noise.
The result does not cover Gaussian noise with a generic correlation kernel.
However, we will see in Section \ref{sec:synthetic-examples} cases that do not fall under the scope of the theorem but still respect the majorization relation when tested numerically.

\begin{figure}[th!]
    \centering
    \includegraphics[width=1.0\textwidth]{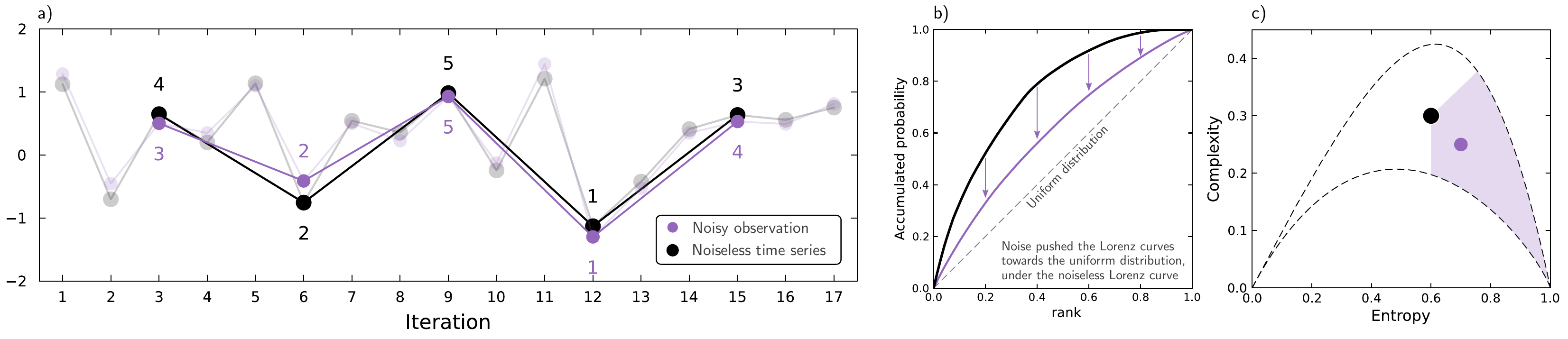}
    \caption{Illustration of how noise on time series reflects in a majorization relationship between ordinal pattern probabilities. (a) The presence of observational noise changes the ordinal pattern for $d  = 5$ and $\tau = 3$ from $\pi = (4,2,5,1,3)$ to $\tilde \pi = (3,2,5,1,4)$. (b) The obtained probabilities $p $ and $\tilde p$ can be represented as Lorenz curves, where now majorization is equivalent to one curve to sit above the other. In this case, the majorization curve associated to $p$ is larger than $\tilde p$. (c) Majorization implies that information-theory metrics also follow a strict order, which allow us to define admissible zones within the complexity-entropy causality planes where noisy ordinal patterns can be found.}
    \label{fig:scheme-majorization}
\end{figure}

Figure \ref{fig:scheme-majorization}a displays an illustration of how majorization affects the probability distribution of the ordinal patterns.
Starting from a noiseless time series, we first compute the value $p(\pi)$ for each possible $D$ permutation $\pi \in S_d$.
Majorization can be represented geometrically by defining the \textit{Lorenz curve} within the box $[0,1] \times [0, 1]$ connecting the points $(k/D, S_k(p))$ for $k = 0, 1, \ldots, D$, with $S_k(p) = \sum_{j=1}^{k} p_j^{\downarrow}$ \cite{lorenz1905methods, Bickel}.
Majorization then implies that the noiseless Lorenz curve associated to $p$ is completely above the Lorenz curve associated to the noisy time series with ordinal pattern distribution given by $\tilde p$, moving the Lorenz curve for $\tilde p$ closer to the uniform distribution characterized by the identity Lorenz curve.
Furthermore, notice that Lorenz curves can be useful for assessing the presence of forbidden patterns, since the number of forbidden patterns corresponds to the ranks at which the cumulative probability has already reached one, so the Lorenz curve stays flat at the top over that range.
A final remark is to notice that ordinal majorization provides a theoretical justification for the probability distributions observed in the multifractal analysis in \textcite{ricci2021permutation-7d6}.

\subsection{Complexity-entropy causality analysis with observational noise}
\label{sec:majorization-to-CE}

The first examples of complexity-entropy causality planes were introduced by \textcite{rosso2007distinguishing-a8b} to classify time-series based on two complementary information-theory metrics: one for entropy and another one for complexity (see \textcite{ribeiro2017characterizing-bfd} for a review on this topic).
The most common metric computed on ordinal patterns is the permutation entropy, defined as
\begin{equation}
    \mathcal{H}[p]
    =
    -\frac{1}{\ln D} \sum_{i=1}^D p_i \ln p_i.
\end{equation}
Permutation entropy by itself is not enough to characterize a time series, which is why complexity metrics have been introduced in the literature \cite{rosso2007distinguishing-a8b}.
Quantifying complexity is a bit more subtle, but common metrics include the Martín–Plastino–Rosso intensive statistical complexity \cite{Martin_Plastino_Rosso_2006} defined as the product of a normalized entropy and a normalized Jensen--Shannon divergence
\begin{equation}
    C[p]
    =
    \mathcal{H}[p] \cdot \mathcal{Q}_{J}[p, u],
\end{equation}
where the normalized disequilibrium is $\mathcal{Q}_{J}[p, u] = {JS[p, u]} \,/\, {JS_{\max}}$, with the Jensen--Shannon divergence between the observed distribution $p$ and the uniform distribution $u = (1/D, \ldots, 1/D)$, and $JS_{\max}$ is the maximum possible value of $JS[p, u]$, used to ensure $\mathcal{Q}_{J} \in [0, 1]$.
An example of complexity-entropy analysis is displayed in Figure \ref{fig:scheme-majorization}c.
For each value of entropy $\mathcal{H}[p]$, there is a minimum and maximum value that the complexity $C[p]$ can take, which determine the minimum and maximum complexity-entropy curves depicted in the same figure \cite{Martin_Plastino_Rosso_2006}.

The link between majorization and the complexity-entropy metrics used in ordinal pattern analysis is via the concept of Schur-convex functions.
A function $G: \R^D \mapsto \R$ is said to be Schur-convex (concave) if $\tilde p \prec p$ implies $G(\tilde p) \leq G(p)$ ($G(p) \leq G(\tilde p)$).
It is a well known fact that the family of functions $G(p) = \sum_{i=1}^D g(p_i)$ with $g$ a convex (concave) function is Schur-convex (concave) \cite{Bickel}.
This implies that both functions $p \mapsto \mathcal{H}(p)$ and $p \mapsto JS[p, u]$ are Schur-concave, and hence we conclude that $\tilde p \prec p$ further implies
\begin{equation}
    H(p) \leq H(\tilde p),
    \qquad
    JS[p, p_e] \leq JS[\tilde p, u],
\end{equation}
and in particular this second inequality further implies
\begin{equation}
    \frac{C[p]}{\mathcal{H}[p]}
    \leq
    \frac{C[\tilde p]}{\mathcal{H}[\tilde p]}.
\end{equation}
In the CE plane, this means that the point $(\mathcal{H}[\tilde p], C[\tilde p])$ associated to $\tilde p$ lies on the right of the noiseless $(\mathcal{H}[p], C[p])$ and under the line traced between the origin and $(\mathcal{H}[p], C[p])$.
This defines an \textit{admissible zone} in the complexity-entropy causality plane where the noisy version of a given time series can exist (see Figure \ref{fig:scheme-majorization}c). 
The presence of a time series in a given location in the complexity-entropy causality plane can then be used to infer where the noiseless version of such time series can be located.

For some applications, entropy and complexity are also not sufficient to fully characterize a time series \cite{guisande2024renyi-fdb}.
Different proposals have been introduced in the literature and we cannot cover them all in this work.
However, it is worth noticing that many of the information-theory metrics used in the literature further correspond to cases of Schur-convex functions, where our analysis can be also applied to defined admissible zones in more generic complexity-entropy causality planes.
Beyond the permutation entropy, more general forms of entropies such as Renyi \cite{guisande2024renyi-fdb}, Tsallis entropies \cite{chen2024tsallis-f5c}, and generalized entropies \cite{Tempesta_2011, Amigo_Dale_Tempesta_2021, amigo2024permutation-8dc} are also used for complexity-entropy analysis.
It is a well known fact that these also are Schur-convex functions.
More recently, the Sharma--Mittal entropy has also been shown to be Schur-concave \cite{Bruno_Vaccaro_2026}, so that the admissible-zone analysis introduced above still holds for this broader family of entropies.

\subsection{Hypothesis Test for Majorization}
\label{sec:hyphotesis-testing}

Although majorization is a well-defined partial ordering between probability vectors, as these vectors need to be estimated with a finite sample of points, we expect errors in the estimates for the probabilities.
This motivates the use of a hypothesis test for evaluating whether a distribution majorizes another.
In this section, we introduce a hypothesis test for testing when a distribution $p$ majorizes another distribution $q$.
We test
\begin{equation}
H_0: p \succ q \quad \text{vs.} \quad H_1: p \not\succ q,
\end{equation}
which is equivalent to testing $S_k(p) \geq S_k(\tilde p)$ for all $k = 1, \ldots, d-1$, with $S_k(p) = \sum_{j=1}^{k} p_j^{\downarrow}$.
Let $\hat p$ and $\hat q$ denote the empirical distributions estimated from independent samples of sizes $n_1$ and $n_2$, respectively.
Let $g_k = S_k(p) - S_k(q)$ denote the $k$-th majorization gap, and let $\hat{g}_k = S_k(\hat{p}) - S_k(\hat{q})$ denote its empirical counterpart.
The sampling variability of each empirical gap $g_k$ can be estimated as the variance of the subtraction of two independent partial sums:
\begin{equation}
\hat{\sigma}_k^2 \;=\; \tfrac{S_k(\hat{p})\,(1 - S_k(\hat{p}))}{n_1} + \tfrac{S_k(\hat{q})\,(1 - S_k(\hat{q}))}{n_2}.
\end{equation}
We can construct a studentization version of the gaps $g_k$ as $\hat z_k = g_k / \max(\hat{\sigma}_k, \sigma_\text{min})$, with $\sigma_\text{min} = 1 / n_\text{eff}$, $n_\text{eff} = n_1n_2 / (n_1 + n_2)$, a minimum allowed variance to avoid division by zero when $S(p), S(q) \approx 0, 1$.
Under $H_0$ we have $z_k \geq 0$ for all $k$.
The test statistic is the most violated constraint defined as
\begin{equation}
\hat T = \min_{1 \leq k \leq d-1 } \, \hat{z}_k,
\end{equation}
with rejection of $H_0$ for sufficiently negative values of $\hat T$.

Since $H_0$ is composite and defined by a system of inequality constraints, the rejection probability is maximized at the least favorable configuration $p = q$.
Calibrating under this configuration yields a valid but conservative test \cite{barrett2003consistent,linton2005consistent}.
To recover power, we adopt the generalized moment selection (GMS) approach of \textcite{andrews2010inference}, which exploits the fact that only the binding constraints ($z_k = 0$) contribute to the asymptotic null distribution of $T$, whereas strictly slack constraints ($z_k > 0$) are asymptotically irrelevant and merely add noise to the bootstrap procedure within the test.
Since the binding/slack status of each constraint is unknown, GMS uses the data to identify which constraints are plausibly binding at the truth.
The selected set of near-binding constraints is then defined as
\begin{equation}
    \hat{\mathcal{K}}
    \;=\;
    \Bigl\{ k \in \{1, \ldots, d-1\} : \hat{z}_k \leq \kappa (n_\text{eff}) \Bigr\}, \qquad \kappa(n) = c \sqrt{\log(n)}.
\end{equation}
An index $k$ is excluded from $\hat{\mathcal{K}}$ when its standardized empirical gap exceeds $\kappa$, which is the empirical signature of a strictly slack constraint at the truth ($g_k > 0$).
The threshold $\kappa(n)$ grows slowly enough that all truly binding constraints are retained ($\kappa(n) / \sqrt{n} \rightarrow 0$), and fast enough that all truly slack constraints are dropped, with probability tending to one ($\kappa(n) \rightarrow \infty$) \cite{andrews2010inference}.
The harmonic-mean effective sample size $n_{\text{eff}}$ replaces the naive $n_1 + n_2$ so that an asymmetric design (e.g.\ a long model simulation against a short data record) does not inflate $\kappa$ and over-select slack constraints.
Bootstrap replicates $(\hat{p}^{*}_b, \hat{q}^{*}_b)$ are drawn by resampling each group independently from its own empirical distribution, equivalently via multinomial draws from $\hat{p}$ and $\hat{q}$.
The recentered bootstrap statistic subtracts the empirical gap from each bootstrap gap, shifting the bootstrap distribution to mimic the null with binding constraints, and takes the minimum only over the selected set:
\begin{equation}
T^{*}_b \;=\; \min_{k \in \hat{\mathcal{K}}} \, \Bigl[ \bigl(S_k(\hat{p}^{*}_b) - S_k(\hat{q}^{*}_b)\bigr) - \hat{g}_k \Bigr].
\end{equation}
The $p$-value is then defined as $B^{-1} \sum_{b=1}^{B} \mathbbm{1}\{T^{*}_b \leq \hat T\}$, and $H_0$ is rejected at level $\alpha$ if the $p$-value is smaller than $\alpha$.
By restricting the minimum to $\hat{\mathcal{K}}$, the bootstrap distribution of $T^{*}_b$ is no longer inflated by clearly-slack constraints, which sharpens the test substantially when only a subset of the $d-1$ majorization constraints are near-binding at the truth \cite{romano2014practical,canay2017practical}.

\begin{figure}[t!]
    \centering
    \includegraphics[width=1.0\textwidth]{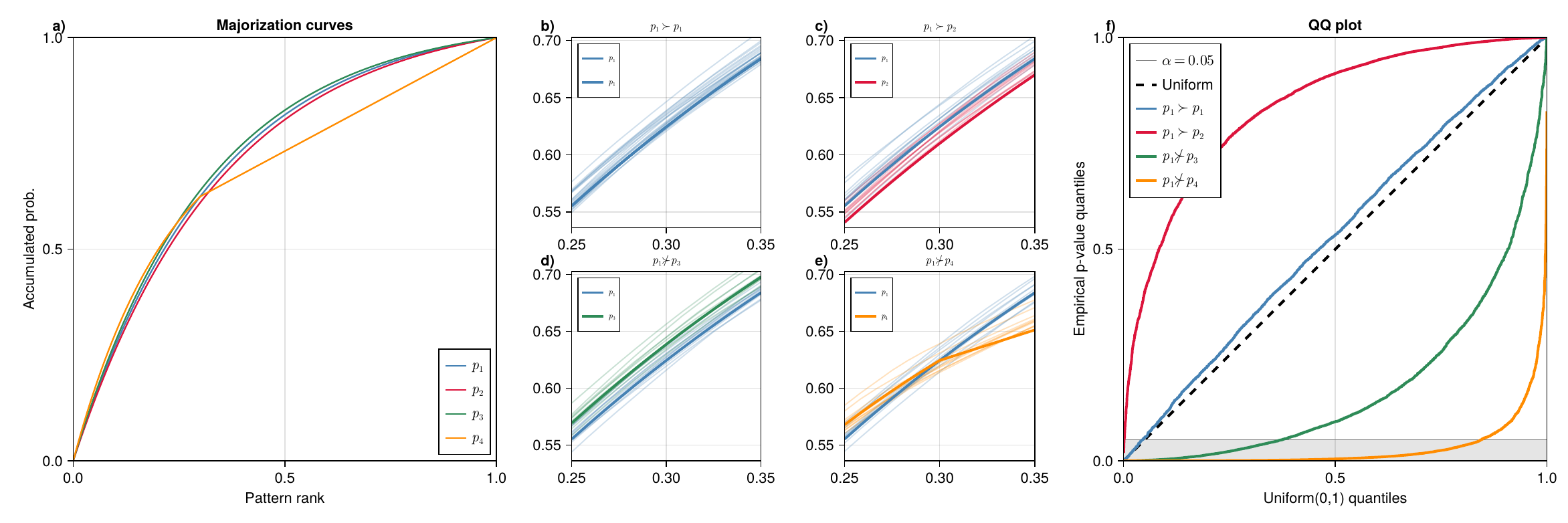}
    \caption{Hypothesis testing for majorization. (a) synthetic Lorenz curves used to generate finite samples. (b-e) Zoom in the majorization curves together with the bootstrap samples associated to $N = 4000$. (f) Distribution of $p$-values for multiple realization of the samples.}
    \label{fig:testing-majorization}
\end{figure}

Figure \ref{fig:testing-majorization} shows how the test performs on a synthetic example.
We consider four Lorenz curves $p_1$ (blue), $p_2$ (red), $p_3$ (green), and $p_4$ (orange) associated to $D = 120$ (Figure \ref{fig:testing-majorization}a).
This example is constructed in such a way that $p_3 \succ p_1 \succ p_2$ and there is no majorization ordering between $p_4$ and $p_1$, $p_2$, and $p_3$.
We consider the four hypothesis test with $H_0: p_1 \succ p_i$ for each $i = 1, 2, 3, 4$ in the finite sample case where probabilities are estimated with a sample size of $N = 4000$.
The bootstrap samples of the curves are shown in sub-panels b-e, showing that statistical dispersion can break the majorization relationship.
The result of the hypothesis test is a unique $p$-value used to reject or fail to reject $H_0$.
We repeat this procedure over independent realizations of the $N$ samples and, in Figure \ref{fig:testing-majorization}f, report the empirical quantiles of the resulting $p$-values against those of a uniform distribution for each of the four tests.
The test $H_0: p_1 \succ p_1$ corresponds to a true null and is used to assess the size of the test; as expected, its $p$-values are uniformly distributed, so the quantiles reproduce the identity line.
The test $H_0: p_1 \succ p_2$ is also a true null, since $p_3 \succ p_1 \succ p_2$ by construction, and accordingly its $p$-values remain above $\alpha = 0.05$ for most realizations, showing that the test does not spuriously reject a genuine majorization relationship.
In contrast, $H_0: p_1 \succ p_3$ and $H_0: p_1 \succ p_4$ are both false nulls, since $p_3$ in fact majorizes $p_1$ and no majorization relation holds between $p_1$ and $p_4$, and the tests correctly reject them, with $p$-values falling below $\alpha = 0.05$ in $\sim 40\%$ and $\sim 80\%$ of realizations respectively, illustrating the power of the test in the finite-sample regime.

\subsubsection{Upper bounds for observational noise}
\label{sec:noise-upper-bound}

Passing the majorization test $p_\text{obs} \prec p_0$ is necessary but not sufficient to conclude that the model with ordinal pattern probabilities $p_0$ correctly describes the observations with probabilities $p_\text{obs}$.
This is because a highly structured model with low entropy will majorize almost any noisy time series, including one generated by an entirely different process.
A more discriminating question is how much observational noise would be required for the model to remain consistent with the observations.
To answer this, we define the \textit{noise upper bound} $\hat\sigma^*$ as follows.
For a given noise level $\sigma \geq 0$, let $p_{\sigma}$ denote the ordinal pattern distribution of the model with added independent Gaussian noise at level $\sigma$.
By Theorem~\ref{theorem:main}, $p_{\sigma_1} \prec p_{\sigma_2}$ for any $\sigma_2 \leq \sigma_1$.
We test, as a function of $\sigma$, whether the noise-corrupted model still majorizes the observations:
\begin{equation}
    H_0(\sigma)\colon p_\text{obs} \prec p_\sigma.
    \label{eq:noise-bound-test}
\end{equation}
For small $\sigma$, $p_{\sigma}$ remains more structured than $p_\text{obs}$ and the test passes.
As $\sigma$ increases beyond the true noise level of the observations, the test fails.
We define
\begin{equation}
    \hat\sigma^* = \min\bigl\{\sigma \geq 0 : H_0(\sigma) \text{ is rejected at level } \alpha\bigr\},
    \label{eq:sigma-star}
\end{equation}
which is an upper bound on the noise level compatible with the reference model explaining the observations under the majorization criterion.
If independent noise estimates for the time series exceed $\hat\sigma^*$, the model cannot explain the observations regardless of any additional calibration or fitting.
An advantage of this approach is that it bounds the noise level without requiring any inference on the original time series, although it still requires specifying the reference noiseless ordinal pattern distribution $p_0$ of the model.
This is a useful property beyond the majorization context, wherever noise estimation without explicit model fitting is needed.
We validate this procedure on synthetic data in Section~\ref{sec:synthetic-examples}.

\section{Results}
In this section, we will first show the performance of our method with a collection of canonical synthetic examples.
We then move our analysis to one of the motivating questions behind this work: to test whether fluctuations in the dipole intensity of the magnetic field are better described by chaotic or stochastic dynamics.
All information-based metrics computed in this work, including complexity and entropies, are computed using the Julia package \texttt{ComplexityMeasures.jl} \cite{ComplexityMeasures.jl}.
Simulation of continuous time series is performed using \texttt{DifferentialEquations.jl} and \texttt{StochasticDiffEq.jl} \cite{DifferentialEquations.jl-2017, rackauckas2017adaptive}.
The synthetic examples used for this work can be found and executed using the following link: \texttt{https://github.com/facusapienza21/NoisyCE.jl}. 

\subsection{Synthetic Examples}
\label{sec:synthetic-examples}

\begin{figure}[th!]
    \centering
    \includegraphics[width=0.9\textwidth]{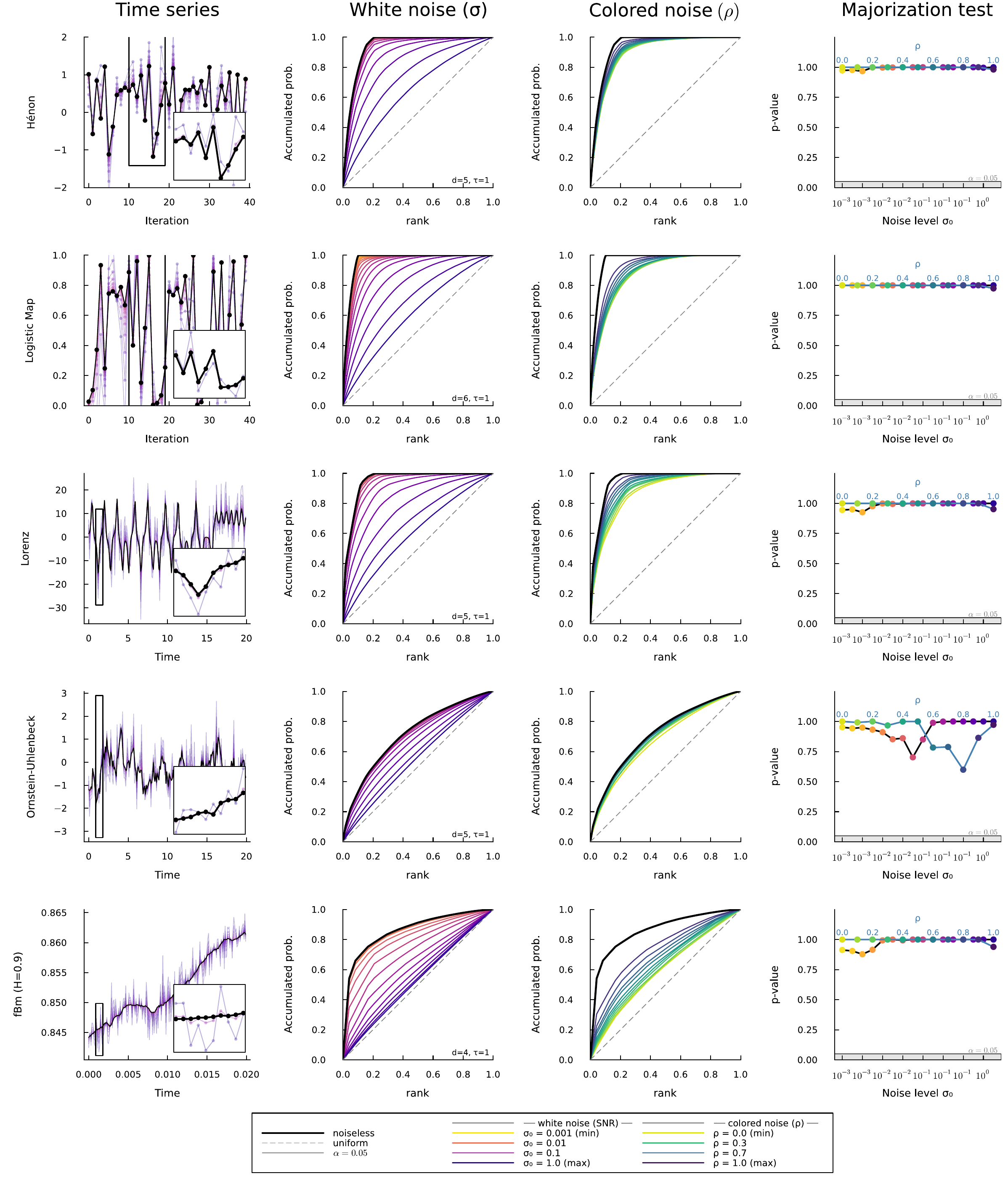}
    \caption{Ordinal majorization analysis for five synthetic benchmarks (rows): Hénon map, logistic map, Lorenz system, Ornstein–Uhlenbeck process, and fractional Brownian motion. First column: representative time series. Second column: Lorenz curves for the noiseless (black) and white-noise-corrupted distributions at increasing levels $\sigma_0$ (colors). Third column: same for colored autoregressive noise at fixed level and varying correlation $\rho$. In both cases, larger noise pushes the Lorenz curve toward the uniform distribution, consistent with $\tilde{p} \prec p$. Fourth column: $p$-values from the majorization hypothesis test (Section \ref{sec:hyphotesis-testing}); dashed line marks $\alpha = 0.05$. A complete specification of the five examples together with the choice of embedding parameters is given in Appendix \ref{sec:appendix-examples}.}
    \label{fig:examples-majorization-1}
\end{figure}

In this section, we illustrate the behavior of ordinal majorization on five synthetic benchmarks, all of them commonly used in the literature: the Hénon map, the logistic map, the Lorenz System, an Ornstein–Uhlenbeck process, and the fractional Brownian motion (see Appendix \ref{sec:appendix-examples} for a complete description of these examples).
We chose these benchmarks to balance both discrete and continuous time series, and chaotic and stochastic time series.
The first column in Figure \ref{fig:examples-majorization-1} shows an example of these five time series.
For each one of these noiseless time series, we can construct the Lorenz curve.
For each time series, we consider the original noiseless distribution of the ordinal patterns together with its associated Lorenz curve and the distorted one obtained by adding white uncorrelated noise $\epsilon_i$ with $\mathbb{E}[\varepsilon_i] = 0$, $\mathbb{E}[\varepsilon_i^2] = \sigma^2$ and $\mathbb{E}[\varepsilon_i \varepsilon_{j}] = 0$ for $i \neq j$ (second column).
In the third column, we repeat the analysis with colored noise of fixed level and variable correlation $\rho$, modeled as an autoregressive process with $\mathbb{E}[\varepsilon_i \varepsilon_{i + k}] = \rho^k \sigma^2$.
The values of $\sigma$ are normalized to represent the noise level $\sigma_0 = \sigma / \sigma_\text{obs}$ (the inverse of the signal-to-noise ratio), with $\sigma_\text{obs}$ the standard deviation of each one of the five synthetic time series (for the fractional Brownian motion, whose variance grows with time rather than remaining fixed, we instead set a fixed noise level directly).
For the colored-noise experiments (third column), the noise level is fixed at $\sigma_0 = 0.2$ while the correlation $\rho$ is varied.
In all the examples shown in Figure \ref{fig:examples-majorization-1}, we observe that larger noise levels move the Lorenz curves closer to the uniform distribution represented by the identity line.
In the case of white noise we are under the conditions of Theorem \ref{theorem:main}, so this majorization relationship is expected to hold.
However, it is interesting to observe that even in the presence of colored noise outside the hypothesis of Theorem \ref{theorem:main}, the majorization relationship still holds in all five examples.
This is further confirmed by our hypothesis test introduced in Section \ref{sec:hyphotesis-testing}, which shows $p$-values above $\alpha = 0.05$ for all test cases (fourth column in Figure \ref{fig:examples-majorization-1}).

\begin{figure}[th!]
    \centering
    \includegraphics[width=0.9\textwidth]{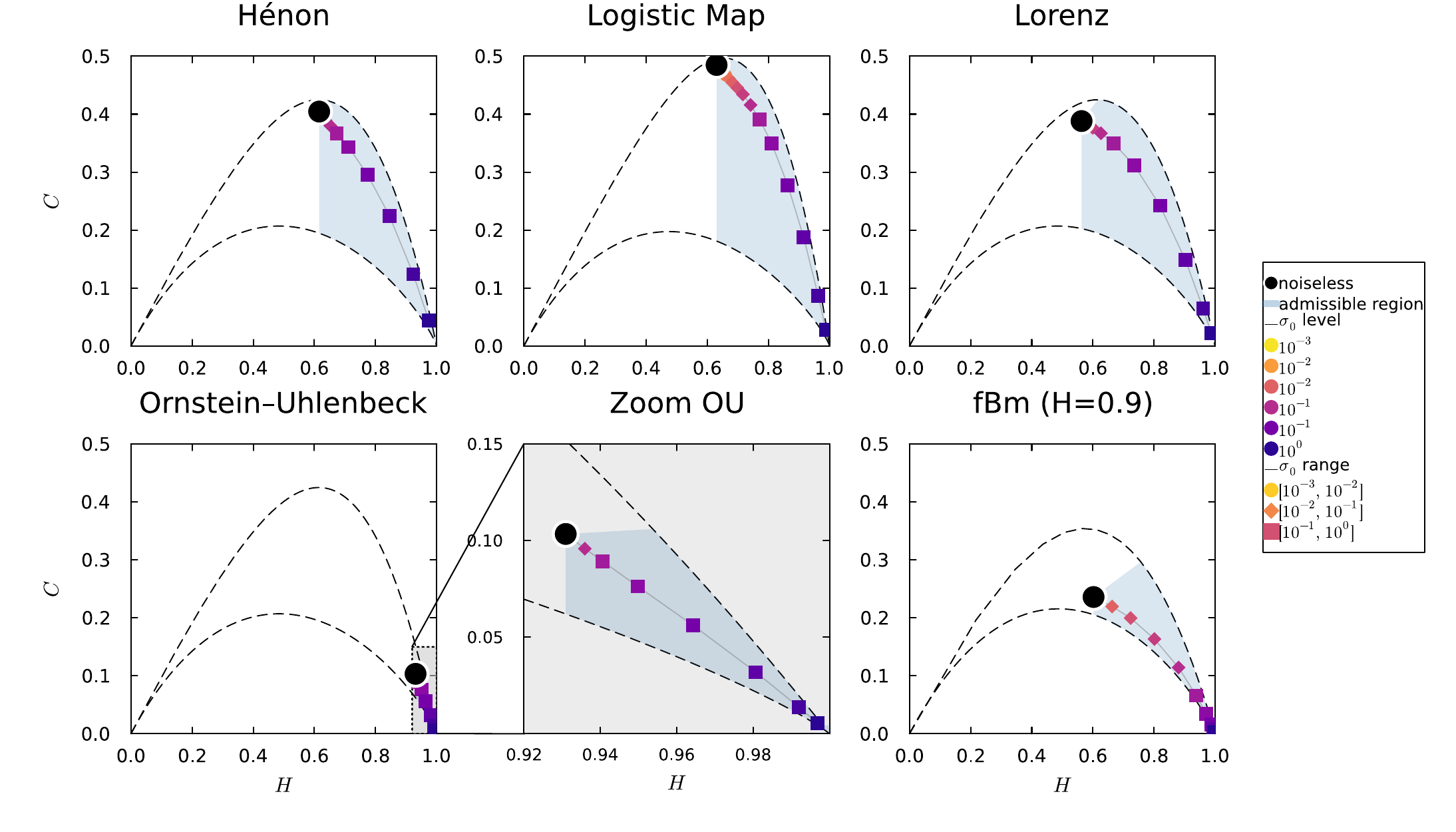}
    \caption{Admissible zones in the complexity-entropy causality plane for each of the five synthetic benchmarks (panels). Each panel shows the noiseless reference point (black circle), the majorization-based admissible zone (blue region), and the trajectory of noisy versions at increasing noise levels $\sigma_0$ (colors). The circle, diamond, and square markers are further used as a visual reference for the different noise levels. In all cases, the noisy time series remains within the admissible zone defined by the noiseless distribution, consistent with the majorization relation established in Section \ref{sec:majorization-to-CE}.}
    \label{fig:examples-majorization-2}
\end{figure}

We now consider the effect of observational noise in the complexity-entropy causality plane.
Figure \ref{fig:examples-majorization-2} shows the complexity-entropy causality plane for each benchmark, together with the majorization-based admissible zone (blue region) introduced in Section \ref{sec:majorization-to-CE}, which is a tighter constraint on where noisy versions can lie than the generic minimum/maximum complexity bounds.
In all five cases, we confirm that the noisy time series lies within the admissible zone defined by each noiseless version and, in particular, by the noiseless reference (black circle in Figure \ref{fig:examples-majorization-2}).
% Figure \ref{fig:examples-CE-noise} further quantifies the variation in entropy and complexity as a function of noise level.
Both metrics remain stable for small values of the noise level $\sigma_0$, but the effect of observational noise becomes significant for $\sigma_0 \approx 0.1$.

\begin{figure}[t!]
    \centering
    \includegraphics[width=1.0\textwidth]{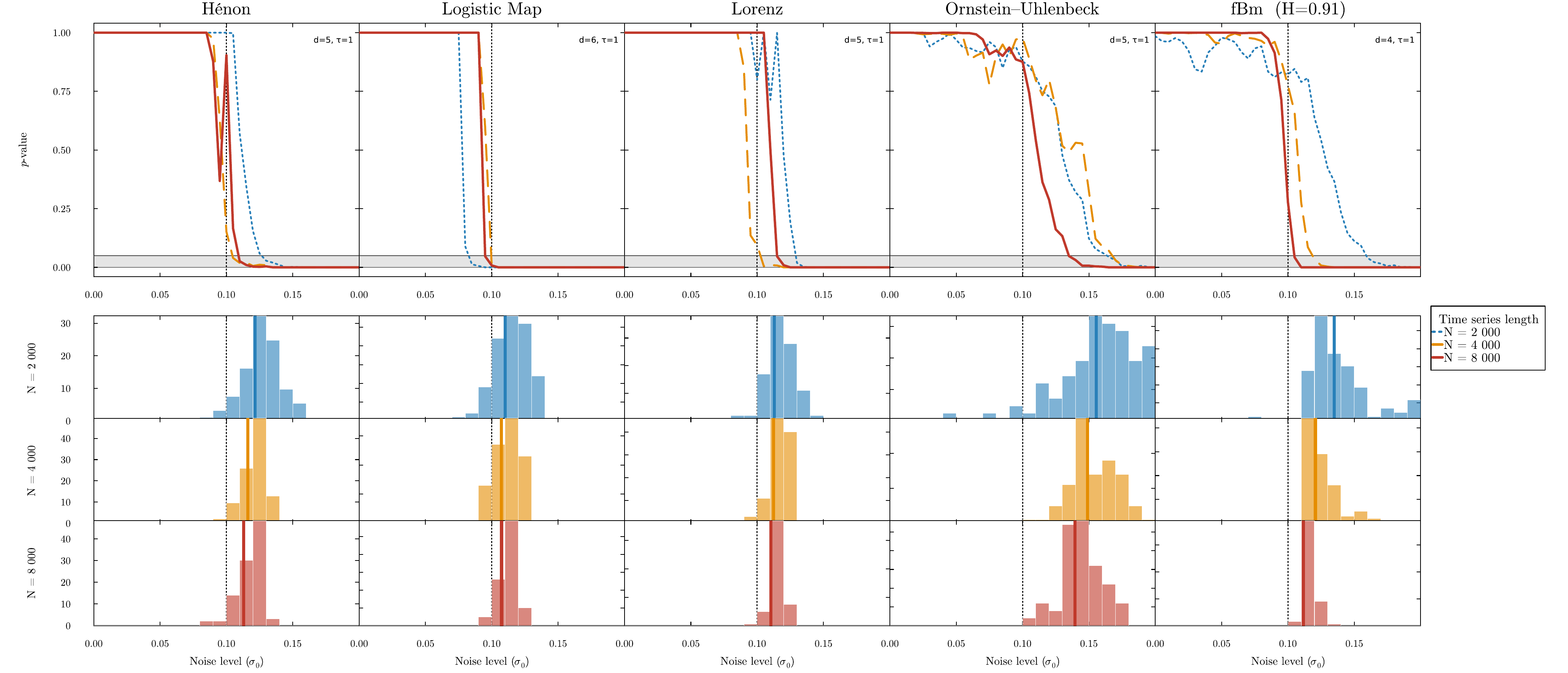}
    \caption{Upper bound for observational noise analysis for the five synthetic benchmarks (columns). First row: $p$-value of the majorization hypothesis test as a function of noise level $\sigma_0$ for a single time series realization, shown for three sample sizes $N = 2000$ (blue), $4000$ (yellow), and $8000$ (red). Second row: distribution of the minimum noise level at which the test fails, estimated over multiple realizations. Vertical lines mark the mean of each distribution. }
    \label{fig:examples-majorization-max-noise}
\end{figure}

Figure~\ref{fig:examples-majorization-max-noise} validates the noise upper bound procedure introduced in Section~\ref{sec:noise-upper-bound} on the same five synthetic benchmarks.
For each benchmark, we generate observations by adding independent Gaussian noise at a known level $\sigma_0 = 0.1$ and estimate $\hat\sigma^*$ by sweeping over $\sigma$ and identifying the first
rejection of $H_0(\sigma)$ (see Equation~\ref{eq:noise-bound-test}).
The first row of Figure~\ref{fig:examples-majorization-max-noise} shows the $p$-value of the majorization test as a function of $\sigma$ for a single realization and three sample sizes $N = 2000, 4000, 8000$.
In all five benchmarks, the $p$-value remains above $\alpha = 0.05$ for $\sigma < \sigma_0$ and drops sharply near $\sigma_0$, confirming that $\hat\sigma^*$ recovers the true noise level.
The second row shows the distribution of $\hat\sigma^*$ over multiple independent realizations of the time series.
We observe that, as $N$ increases, the distribution tightens and its mean converges to $\sigma_0$.
This demonstrates that the upper bound can be estimated reliably without any parametric inference or explicit noise model fitting, using only the hypothesis test introduced in Section~\ref{sec:hyphotesis-testing}.

\subsection{Application to Paleomagnetism and the Dynamic of Earth Magnetic Dipole}
\label{sec:real-examples}

\begin{figure}[th!]
    \centering
    \includegraphics[width=1.0\textwidth]{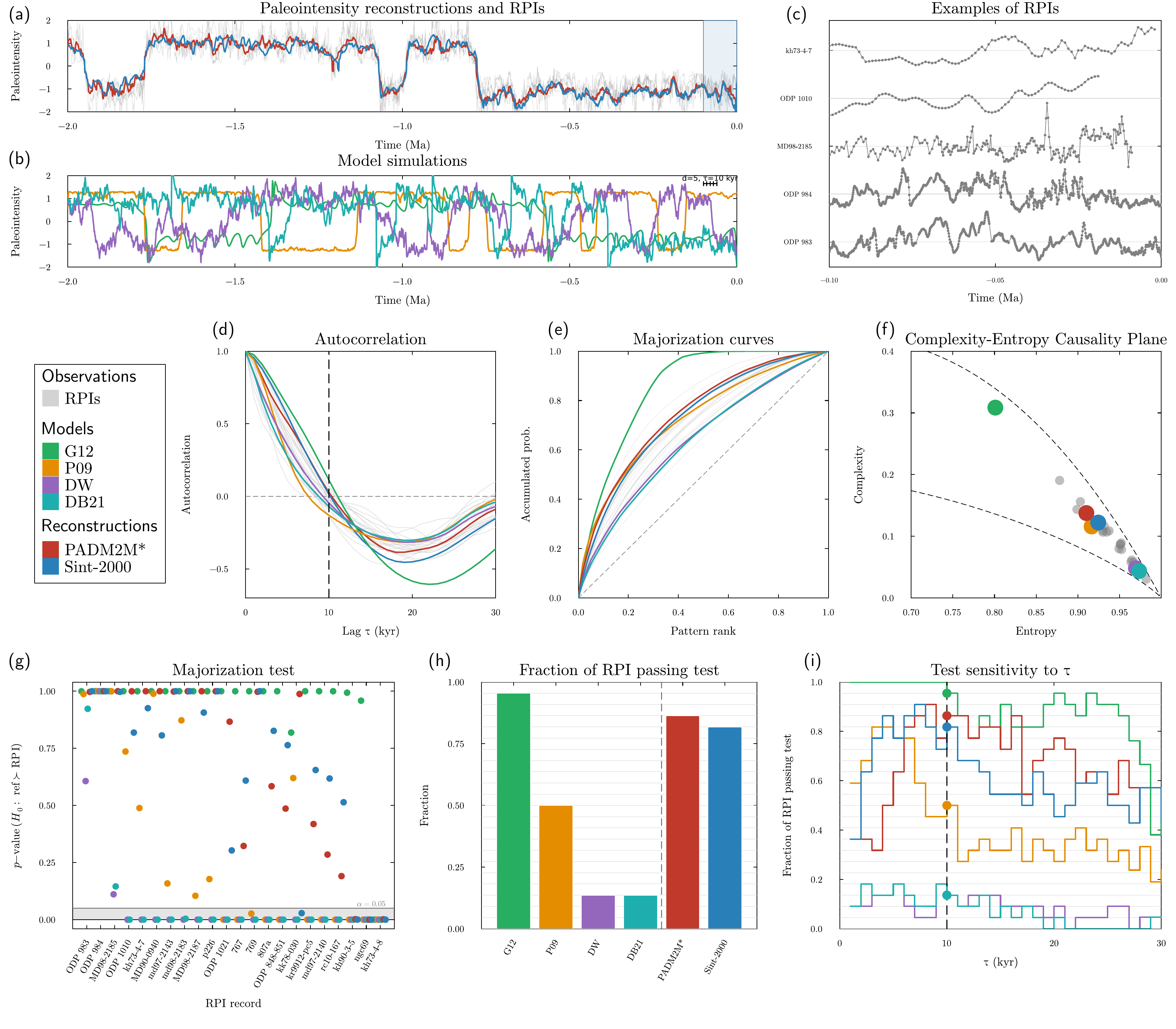}
    \caption{Summary of the ordinal majorization analysis applied to geomagnetic analysis. (a) Paleointensity reconstructions PADM2M* (red) and Sint-2000 (blue) together with the examples of filtered RPI records (gray) over the past 2 Ma; the shaded window marks the zoom region shown in (c). (b) Example simulations from the four reduced-order models G12 (green), P09 (orange), DW (purple), DB21 (turquoise); the scale bar indicates the embedding window ($d = 5$, $\tau = 10$ kyr). (c) Zoom-in on five individual RPI records in the highlighted window, normalized and offset vertically. (d) Autocorrelation functions for all models, reconstructions, and RPI records (gray); the dashed vertical line marks the selected $\tau = 10$ kyr. (e) Lorenz curves for models and reconstructions (colors) and RPI records (gray). (f) Complexity-entropy causality plane; gray circles are RPI records, colored circles are models and reconstructions. (g) $p$-values from the majorization hypothesis test for each of the 22 records and all 6 references; shaded region and dashed line mark $\alpha = 0.05$. (h) Fraction of RPI records passing the majorization test for each reference. (i) Fraction passing as a function of embedding delay $\tau$; dashed line marks $\tau = 10$ kyr.}
    \label{fig:examples-majorization-geomagnetic-1}
\end{figure}

The dynamics of the Earth's time-varying magnetic field, known as the geodynamo, is one of the most fascinating yet challenging physical processes in geophysics.
Fluctuations of the geodynamo occur at a broad range of timescales, exhibiting short term variations to long-term changes such as the reversal of the dominant axial dipole \cite{Courtillot_Mouel_1988, Sadhasivan_Constable_2022}.
During a reversal, the intensity of the magnetic field weakens in prelude to a change in direction of the dipole component of Earth's magnetic field (see Figure \ref{fig:examples-majorization-geomagnetic-1}a).
Understanding the dynamics of geomagnetic reversals is key to understanding the physical processes underlying Earth's dynamo dynamics and planetary behavior and have important impacts on Earth's biosphere \cite{Channell_Vigliotti_2019, Pan_Li_2023}, and modern technological infrastructure including power grids and satellite systems \cite{Oughton_2019}, among others. 
For a good review on these topics, we direct the interested reader to \textcite{Hulot_Finlay_Constable_Olsen_Mandea_2010}. 

Modelling how Earth's conducting liquid outer core generates a sustained geodynamo requires solving complex three-dimensional magneto-hydrodynamics partial differential equations, and can be computationally very demanding \cite{glatzmaiers1995three, Schaeffer_Jault_Nataf_Fournier_2017}, motivating the development of reduced-order models that capture key properties of Earth's dipole. 
These models fall into two families: those based on low-dimensional chaos and those based on scalar stochastic differential equations \cite{Gwirtz_Morzfeld_Fournier_Hulot_2020}. 
The former represent deterministic reduced models that reproduce the random-like behavior of geomagnetic reversals in good agreement with the turbulent nature of the equations used to describe the dynamo.
For example, the G12 model \cite{Gissinger_2012} introduces a simple system of three ODEs.
On the other hand, models based on stochastic differential equations include P09 \cite{Petrelis_Fauve_Dormy_Valet_2009}, the double-well (DW) potential SDE model \cite{Morzfeld_Buffett_2019}, and SDEs with auto-correlated noise DB21 \cite{Davis_Buffett_2021}.
Figure \ref{fig:examples-majorization-geomagnetic-1}b shows examples of one simulation for each one of these four models. 
Both families of models are widely used in the literature to study and simulate the random-like behavior of Earth's dipole (e.g., \textcite{Gwirtz_Morzfeld_Fournier_Hulot_2020}, \textcite{Gwirtz_Davis_Morzfeld_Constable_Fournier_Hulot_2022}, \textcite{Sadhasivan_Constable_2022}).
A natural question arises: \textit{Does the evidence based on paleomagnetic data support a model for Earth's dipole based on chaotic or stochastic dynamics?}

\subsubsection{Paleomagnetic observations}

In order to address this question, we consider a dataset of 86 relative paleointensity (RPI) records derived from marine sediment cores distributed globally spanning the last 4 Ma (Figure \ref{fig:examples-majorization-geomagnetic-1}c).
These observations can be used to recover absolute paleointensities by understanding the physical and chemical processes involved in how the marine sediments acquired a remanent magnetization recording Earth's ancient magnetic field \cite{Valet_2003}.
For this work we consider the same RPIs compiled in \textcite{Ziegler_Constable_Johnson_Tauxe_2011} which are archived in the EarthRef Digital Archive (\texttt{https://earthref.org/ERDA/1139/}).
A subset of the RPIs sequences is shown in Figure \ref{fig:examples-majorization-geomagnetic-1}c.
RPIs are further converted to signed paleointensities using the geomagnetic polarity time scale GPTS-CK95 \cite{cande1995revised}, which provides sub-kyr resolution of reversal timing over the past 5 Ma. 

Ordinal pattern analysis is particularly well-suited for this problem for two reasons.
First, the sequential deposition of sediments provides a naturally ordered time series.
Second, since RPIs are only a proxy for the actual dipole intensity --- with a linear (or sometimes non-linear \cite{Ziegler_Constable_Johnson_Tauxe_2011}) dependence mediated by the amount and type of magnetic minerals in the sediment and environmental factors --- a direct calibration to absolute intensities is required in most analyses.
However, since monotone transformations do not affect ordinal pattern statistics, we can analyze the RPIs directly without such calibration, which remains an area of active research \cite{Ziegler_Constable_Johnson_Tauxe_2011}.
We select only the subset of 22 marine records for which $N \geq 5D$, with $d = 5$, following the recommendation in \textcite{Riedl_Muller_Wessel_2013, amigo2008combinatorial}, and covering a temporal span of at least $0.2$ Ma.
We further consider two paleomagnetic reconstructions datasets for analysis, the PADM2M \cite{Ziegler_Constable_Johnson_Tauxe_2011} and Sint-2000 \cite{Valet_Meynadier_Guyodo_2005} datasets.

An important caveat is that RPIs are not uniformly sampled in time, which can affect ordinal pattern statistics.
We adopt the strategy of resampling each RPI record to a uniform time grid via linear interpolation, although this can in principle introduce biases in specific ordinal patterns.
A more general and principled alternative is to resample the reference model simulations with the same irregular time spacing as the observations, which avoids any interpolation and applies naturally to poorly sampled records.
In practice, we find that both approaches yield consistent results for $\tau > 4$ kyr --- larger than the mean ($0.54$ kyr), $0.95$th ($1.9$ kyr), and $0.99$th ($3.68$ kyr) percentiles of the gap distribution --- as the effect of irregular sampling becomes negligible once $\tau$ exceeds the typical inter-sample gaps. 

\subsubsection{Ordinal majorization of geomagnetic models}

We now address the question of whether the observations resulting from the RPIs can be described as noisy observations of the four models previously introduced. 
We first proceed with the selection of the embedding delay $\tau$. 
Following \textcite{Rosenstein_Collins_Luca_1994} and \textcite{Micco_Fernandez_Larrondo_Plastino_Rosso_2012}, we select $\tau$ as the first zero crossing of the autocorrelation function. 
Based on the results shown in Figure \ref{fig:examples-majorization-geomagnetic-1}d for the 4 models, 2 reconstructions, and 22 selected RPIs, we select $\tau = 10$ kyr.

Once $\tau$ is selected, we can compute the probabilities associated with each ordinal pattern. 
Figure \ref{fig:examples-majorization-geomagnetic-1}e shows the majorization curves associated with the four models, each color representing a different model.
The gray lines are the majorization curves corresponding to each one of the RPI records.
We can see the chaotic G12 model sits above the other majorization curves, while the stochastic models are closer to the uniform distribution. 
This result can also be confirmed by looking at the complexity-entropy causality plane (Figure \ref{fig:examples-majorization-geomagnetic-1}f), where we observe that the G12 model is in the admissible zone of all the RPI records. 
In order to quantify the majorization results, we applied the majorization hypothesis test to each pair of model and RPI record, testing whether the noiseless model majorizes the noisy RPI time series. 
Figure \ref{fig:examples-majorization-geomagnetic-1}g shows the $p$-value associated with each one of the 22 selected marine records, and Figure \ref{fig:examples-majorization-geomagnetic-1}h shows the fraction of RPI records that passed the statistical test.
We find that only the G12 model is consistent with the paleomagnetic observations under the majorization criterion. 
As has been emphasized in the literature, we also perform a multi-scale analysis to study the robustness of these results to the choice of $\tau$.
Figure \ref{fig:examples-majorization-geomagnetic-1}i shows the fraction of passed tests as a function of $\tau$, confirming that the chaotic G12 model represents the best model candidate based on the ordinal majorization criteria introduced in this work.

\subsubsection{Final remarks}

The ordinal majorization analysis provides strong evidence that the dynamics of Earth's geomagnetic dipole are better described by the low-dimensional chaotic model G12 than by any of the other stochastic models considered here, which is in good agreement with the results reported in the literature (e.g., \textcite{Ryan_Sarson_2008}, \textcite{Gwirtz_Morzfeld_Fournier_Hulot_2020}).
Crucially, this conclusion is reached directly from the raw RPI records, without any smoothing, calibration, or parametric fitting of the observations: the majorization criterion requires only the ordinal structure of the time series, making the result robust to the amplitude ambiguities inherent in paleointensity proxies.
We also find that the two paleomagnetic reconstructions PADM2M and Sint-2000 are consistent with the individual RPI records under the majorization criterion, confirming that these stacks preserve the ordinal pattern structure of the underlying observations.
This is a non-trivial result, since stack averaging can over-smooth the original series and distort its dynamical character.

The failure of the stochastic models should not be interpreted as evidence that stochastic dynamics are fundamentally incompatible with geomagnetic reversals.
Rather, it reflects the fact that none of the specific models tested here reproduce the ordinal pattern distribution of the observations.
A well-known pathway to reconcile stochastic and chaotic behavior is to introduce long-range temporal correlations in the noise term, which shifts the ordinal pattern distribution toward areas of the complexity-entropy causality plane with higher complexity \cite{rosso2007distinguishing-a8b}.
This observation provides direct motivation for developing stochastic models with long-memory structure as an alternative family of candidates for the geodynamo.

Finally, while we have focused on four representative reduced-order models, the majorization framework extends naturally to any candidate model of the geodynamo.
Beyond model selection, the hypothesis test introduced in Section~\ref{sec:hyphotesis-testing} can serve as a tool for model refinement: the noise upper bound $\hat\sigma^*$ (Section~\ref{sec:noise-upper-bound}) provides a quantitative constraint on how much observational noise a given model can tolerate while remaining consistent with the data, offering a principled criterion for comparing and improving reduced-order models.

\section{Conclusions}
We have shown that observational noise induces a majorization ordering on the ordinal pattern distributions of time series.
Under certain conditions on the noise structure which include the commonly used independent additive noise model, the noiseless distribution majorizes the noisy one, regardless of whether the underlying dynamics are chaotic or stochastic.
This result connects directly to the complexity-entropy causality plane, where majorization defines an admissible zone that constrains where a noisy time series can be located relative to its noiseless counterpart.
Together, these results extend the ordinal pattern framework beyond the small-noise regime where previous results were confined, providing a principled basis for dynamical classification under realistic measurement conditions.

To make this framework operational, we introduced two complementary tools.
First, a hypothesis test for majorization based on the generalized moment selection bootstrap, which accounts for finite-sample variability and remains valid when only a subset of the majorization constraints are binding.
Second, a noise upper bound estimator that identifies the maximum noise level under which a reference model remains consistent with the observations, without requiring any parametric inference or model fitting on the original time series.
This second tool has broad applicability beyond the present context as it provides a model-free way to characterize observational noise in any setting where a candidate noiseless reference is available.

Applied to paleomagnetic records, the framework provides strong evidence that the dynamics of Earth's geomagnetic dipole are better described by a low-dimensional chaotic model than by any of the scalar stochastic models tested.
This conclusion is reached directly from sparse, irregularly sampled relative paleointensity records, without calibration or smoothing, exploiting the invariance of ordinal statistics to monotone transformations.
While the stochastic models tested here fail the majorization criterion, this points toward the development of stochastic models with long-range temporal correlations as a promising direction for future geodynamo modeling.

\section{Perspectives}
Several theoretical extensions of this work remain open.
The conditions of Theorem~\ref{theorem:main} cover independent additive noise and equicorrelated Gaussian noise, but more general correlation structures, including a Gaussian process with a generic kernel, fall outside its scope, even though the majorization relation is observed to hold numerically in all examples tested.
A natural conjecture is that, in the small-noise regime, an approximate or relaxed form of majorization can be established for a broader class of noise models.
Beyond the specific noise conditions, the admissible zone construction can be extended to other information-theoretic metric that is a Schur-convex function of the ordinal pattern distribution, which includes Rényi, Tsallis, and other generalized entropies. Characterizing exactly which complexity-entropy causality planes admit such zones is a direction worth pursuing.

On the applied side, we hope the methodology introduced here finds use in other fields where time series are sparse, noisy, or only available as proxies for the quantity of interest.
Natural candidates include neuroscience, where spike train recordings are short and contaminated by measurement noise; climate science, where proxy records such as ice cores and tree rings share many of the characteristics of paleomagnetic data; and financial time series, where low signal-to-noise ratios are common.
More immediately, the failure of the scalar stochastic models in the geomagnetic analysis motivates the development of reduced-order models with long-range temporal correlations, for which the majorization test provides a concrete and model-free criterion for validation.

\section{Acknowledgments}
I thank Dr. Kyle Gwirtz of NASA’s Goddard Space Flight Center for helpful discussions, feedback, and help with the manipulation of the paleomagnetic datasets.
I also thank Dr. Pablo Groisman and Gabriel Mindlin from the University of Buenos Aires for useful discussions and feedback. 
I thank Margaret Doyle for English edits.
The author used Claude Sonnet 4.6 (Anthropic) to assist with editing the manuscript, including improvements to English grammar and writing clarity, and coding, primarily dedicated to figure generation and testing purposes.

\begin{appendix}
\section{Proof of Theorem}
\label{sec:appendix-proof}
\begin{proof}
For a given values of $d$ and $\tau$, let us define $x_t^{(d,\tau)} = (x_t, x_{t+\tau}, \ldots, x_{t + (d-1)\tau})$ and $y_t^{(d,\tau)} = (y_t, y_{t+\tau}, \ldots, y_{t + (d-1)\tau})$, with $y_t = x_t + \varepsilon_t$.
We construct a stochastic matrix $T \in \mathbb{R}^{D \times D}$ indexed by pairs $(\pi, \widetilde{\pi}) \in S_d \times S_d$, where
\begin{equation}
    T(\pi \to \widetilde{\pi}) = \mathbb{P}\left(
        y_t^{(d,\tau)} \text{ has ordinal pattern } \widetilde{\pi}
        \, \Big | \,
        x_t^{(d,\tau)} \text{ has ordinal pattern } \pi
    \right).
\end{equation}
By strict stationarity of $\{x_t\}$ and the time-invariance of the noise model, this conditional probability is well defined as it does not depend on $t$.
By absolute continuity, ties occur with probability zero in both ${x}_t^{(d,\tau)}$ and ${y}_t^{(d,\tau)}$, so the ordinal pattern is defined almost surely.
By construction, $T$ is a row-stochastic matrix, that is $\sum_{\widetilde{\pi} \in S_d} T(\pi \to \widetilde{\pi}) = 1$ for all $\pi \in S_d$, and we have
\begin{equation}
    \widetilde{p}(\widetilde{\pi})
    = \sum_{\pi \in S_d} T(\pi \to \widetilde{\pi})\, p(\pi),
    \qquad \forall\, \widetilde{\pi} \in S_d,
\end{equation}
or in matrix form $\widetilde{p} = T^\top p$.

Let us now show that $T$ is doubly stochastic, i.e.\ that $\sum_{\pi \in S_d} T(\pi \to \widetilde{\pi}) = 1$ for all $\widetilde{\pi} \in S_d$.
We claim that $T(\pi \to \widetilde{\pi})$ depends only on the relative permutation $\widetilde{\pi} \circ \pi^{-1}$, and not on $\pi$ itself.
Furthermore, we will show that $T(\pi \to \tilde \pi) = T(\{ 1, 2, \ldots, d\} \to \tilde \pi \circ \pi^{-1})$.
In order to prove that, notice
\begin{equation}
    % \mathbb{P}(\pi \to \tilde \pi)
    T(\pi \to \tilde \pi)
    =
    \mathbb{P}\left( y_{\tilde \pi(j)} - y_{\tilde \pi(i)} \geq 0,\; \forall i < j \,\big|\, \pi \right),
\end{equation}
which by conditioning on $\Delta$ becomes
\begin{equation}
    % \mathbb{P}(\pi \to \tilde \pi)
    T(\pi \to \tilde \pi)
    =
    \int \mathbb{P}\left( y_{\tilde \pi(j)} - y_{\tilde \pi(i)} \geq 0,\; \forall i < j \,\big|\, \Delta,  \pi \right)
    \mathbb{P}(\Delta \mid \pi)\, d\Delta.
\end{equation}
Here $\Delta = (\Delta_1, \ldots, \Delta_{d-1})$
% $\Delta = \{\Delta_{i,j}\}_{i \leq j}$ 
denotes the unsigned gaps indexed by rank, i.e.\ 
$\Delta_{i} = x_{\pi(i+1)} - x_{\pi(i)} \geq 0$ for $i = 1, \ldots d-1$, 
% $\Delta_{i,j} = x_{\pi(j)} - x_{\pi(i)} > 0$ for $i < j$, 
so that $\Delta$ is a function of the sorted values of ${x}_t^{(d,\tau)}$ alone.
Next, the integrand can be evaluated as
\begin{equation}
    \mathbb{P}\left( y_{\tilde \pi(j)} - y_{\tilde \pi(i)} \geq 0,\; \forall i < j \,\big|\, \Delta,  \pi \right)
    =
    \mathbb{P}\left( x_{\tilde \pi(j)} - x_{\tilde \pi(i)} \geq \varepsilon_{\tilde \pi(i)} - \varepsilon_{\tilde \pi(j)},\; \forall i < j \,\big|\, \Delta,  \pi \right),
\end{equation}
which using the rank-indexed gap definition becomes
\begin{equation}
    \mathbb{P}\left( y_{\tilde \pi(j)} - y_{\tilde \pi(i)} \geq 0,\; \forall i < j \,\big|\, \Delta,  \pi \right)
    =
    \mathbb{P}\left( \mathrm{sg}(\sigma(i,j))\,\Delta_{\sigma(i), \sigma(j)} \geq \varepsilon_{\tilde \pi(i)} - \varepsilon_{\tilde \pi(j)},\; \forall i < j \,\big|\, \Delta,  \pi \right),
\end{equation}
where we introduced the notation $\Delta_{i,j} = x_{\pi(j)} - x_{\pi(i)} = \Delta_i + \Delta_{i+1} + \ldots + \Delta_{j-1}$ for $i < j$ to denote the positive gap between the sorted values of the embedding.
By the conditional permutation-invariance of $\varepsilon$ given $(\Delta, \pi)$, we may relabel the indices of $\varepsilon$ via $i \mapsto \tilde\pi(i)$ to obtain
\begin{equation}
    \mathbb{P}\left( y_{\tilde \pi(j)} - y_{\tilde \pi(i)} \geq 0,\; \forall i < j \,\big|\, \Delta,  \pi \right)
    =
    \mathbb{P}\left( \mathrm{sg}(\sigma(i,j))\,\Delta_{\sigma(i), \sigma(j)} \geq \varepsilon_i - \varepsilon_j,\; \forall i < j \,\big|\, \Delta, \pi \right)
    =: F_\sigma(\Delta, \pi),
\end{equation}
where $\sigma := \tilde \pi \circ \pi^{-1}$ and $\mathrm{sg}(\sigma(i,j)) := \mathrm{sgn}(\sigma(j) - \sigma(i))$.
The conditional law of $\varepsilon$ given $(\Delta, \pi)$ is permutation-invariant by assumption, so it depends on $\pi$ at most through the symmetric structure preserved by every permutation; in particular, the joint law of the difference vector $(\varepsilon_i - \varepsilon_j)_{i<j}$ given $(\Delta, \pi)$ is fully exchangeable in its component indices and so is determined by $\Delta$ alone in terms of how it pairs with the inequalities indexed by $\sigma$.
Hence $F_\sigma(\Delta, \pi) = F_\sigma(\Delta)$ depends on $(\pi, \tilde\pi)$ only through $\sigma$.

Substituting back,
\begin{equation}
    T(\pi \to \tilde\pi) = \int F_\sigma(\Delta)\, \mathbb{P}(\Delta \mid \pi)\, d\Delta.
\end{equation}
By the assumption that the distribution of the unsigned gaps $\Delta$ is independent of the permutation order, $\mathbb{P}(\Delta \mid \pi) = \mathbb{P}(\Delta)$, and therefore
\begin{equation}
    T(\pi \to \tilde\pi) = \int F_\sigma(\Delta)\, \mathbb{P}(\Delta)\, d\Delta =: \widehat T(\sigma),
\end{equation}
which depends only on $\sigma = \tilde\pi \circ \pi^{-1}$.
Taking $\pi = \mathrm{id}$ yields $\widehat T(\sigma) = T(\mathrm{id} \to \sigma)$, so
\begin{equation}
    T(\pi \to \tilde\pi) = T(\mathrm{id} \to \tilde\pi \circ \pi^{-1}),
\end{equation}
as claimed.

We can now verify that $T$ is doubly stochastic.
For any $\tilde\pi \in S_d$,
\begin{equation}
    \sum_{\pi \in S_d} T(\pi \to \tilde\pi)
    = \sum_{\pi \in S_d} T(\mathrm{id} \to \tilde\pi \circ \pi^{-1})
    = \sum_{\sigma \in S_d} T(\mathrm{id} \to \sigma)
    = 1,
\end{equation}
where the second equality uses the bijection $\pi \mapsto \tilde\pi \circ \pi^{-1}$ on $S_d$, and the third equality is row-stochasticity applied to the row $\pi = \mathrm{id}$.
Hence $T$ is doubly stochastic.
Since $\widetilde p = T^\top p$ with $T^\top$ doubly stochastic, the Hardy--Littlewood--P\'olya theorem yields $\widetilde p \prec p$ \cite{hardy1929inequalities, Bickel}.
\end{proof}

\section{Synthetic Examples}
\label{sec:appendix-examples}
\noindent\textbf{Hénon Map.}
The Hénon map is a discrete 2D chaotic time series introduced in \textcite{Henon_1976}.
It is a canonical low-dimensional chaotic map commonly used for benchmarking (e.g., \textcite{rosso2007distinguishing-a8b}).   
It is described by the recurrence 
\begin{align}
    x_{n+1} &= 1 - a x_n^2 + y_n \\
    y_{n+1} &= b x_n
\end{align}
with parameters $a = 1.4$, $b = 0.3$ and initial condition $x_0 = 0, y_0 = 0$. 
We use the $x$-component after discarding the first $10^4$ transient iterates.
For this experiment we select an embedding dimension $d = 5$ and embedding delay $\tau = 1$ (as commonly used for discrete time series).

\vspace{1mm}
\noindent\textbf{Logistic Map.}
For a second demonstration of discrete chaotic system, we consider the logistic map defined as 
\begin{equation}
    x_{n+1} = r x_n (1 - x_n)
\end{equation}
with initial condition $x_0 = 0.1$ and parameter $r = 4$. 
The logistic map is interesting to analyze as it has forbidden patterns that are very sensitive to observational noise (e.g., \textcite{Amigo_Kocarev_Szczepanski_2006} and \textcite{amigo2010permutation}). For this example we use $d = 6$.

\vspace{1mm}
\noindent\textbf{Lorenz System.}
We consider the $x$-component of the classic Lorenz attractor, governed by the system of ordinary differential equations:
\begin{align}
    \frac{dx}{dt} & = \sigma(y - x) \\
    \frac{dy}{dt} &= x(\rho - z) - y \\
    \frac{dz}{dt} &= xy - \beta z 
\end{align}
with  parameters $(\sigma, \rho, \beta) = (10, 28, 8/3)$ and initial condition $(x,y,t)(0) = (1,1,1)$.
Numerical integration is carried with the Tsitouras solver implemented within \texttt{DifferentialEquations.jl} with small numerical tolerances.
We discard the first $100$ time units of observations. 
The choice of the embedding delay $\tau$ is more delicate for continuous time series.
Following \textcite{Micco_Fernandez_Larrondo_Plastino_Rosso_2012}, we sweep the range value of $\tau$ and pick the value that maximized the complexity metric. 
% Figure \ref{fig:lorenz-delay} shows the values of complexity-entropy as a function of the delay embedding $\tau$.
For simulations in the main text, we selected $\tau = 0.1$, which is in good correspondence with values of the delay used in the literature. 

% \begin{figure}[th!]
%     \centering
%     \includegraphics[width=0.5\textwidth]{figures/delay_lorenz_CH.png}
%     \caption{...}
%     \label{fig:lorenz-delay}
% \end{figure}

\vspace{1mm}
\noindent\textbf{Ornstein–Uhlenbeck Process.}
We consider a dynamical system driven by a stochastic differential equation describing the Ornstein–Uhlenbeck Process:
\begin{equation}
    dX_t = - \theta (X_t - \mu) + \sigma dW_t,
\end{equation}
with $\mu$ the mean of the process, and $\sigma$ and $\theta$ controlling the aleatoric noise level. 
The stationary distribution of the process is a Gaussian with mean $\mu$ and standard deviation $\sigma$. 
The autocorrelation of the time series is $\mathbb{E}[(X_t - \mu)(X_s - \mu)] = \sigma^2 \exp(- |t - s| / \theta)$, so $\theta$ control the correlation length. 
Numerical integration is not required as the OU process has an exact discrete update rule \cite{gillespie1996exact}. 
For our simulations, we simply do $(\mu, \theta, \sigma) = (0,1,1)$.
Unlike the chaotic examples, the OU process is inherently stochastic, so its ordinal distribution already carries randomness.
Nevertheless it is still structured (non-uniform) due to temporal correlations for $\theta > 0$.
As we want to capture the correlation structure of the time series, we pick $\tau = \theta$, which is the natural length scale of the stochastic process.

\vspace{1mm}
\noindent\textbf{Fractional Brownian Motion.}
Fractional Gaussian noise (fGn) is a zero-mean Gaussian process fully characterized by the auto-covariance function
\begin{equation}
  C_H(s) = \tfrac{1}{2}\bigl(|s+1|^{2H} - 2|s|^{2H} + |s-1|^{2H}\bigr),
\end{equation}
where $H \in (0,1)$ is the Hurst exponent \cite{mandelbrot1968}. 
The process is anti-persistent for $H<1/2$, recovers white Gaussian noise at $H=1/2$, and is long-range correlated for $H>1/2$, with $C_H(s) \sim
s^{2H-2}$ as $s \to \infty$.
For simulations in the main text we used $H = 0.9$.
It is interesting to remark that the fGn is non-stationary and then it does not fall under the conditions of Theorem \ref{theorem:main}, although numerically we observe the result still applies.

\end{appendix}

% \newpage
\printbibliography[heading=bibintoc, title={References}]

\end{document}